\documentclass[aps,prl,twocolumn,showpacs,superscriptaddress,floatfix,nofootinbib]{revtex4}
\usepackage{graphicx}
\usepackage{amsmath}
\usepackage{epsfig}
\usepackage{helvet}
\usepackage{amssymb}

\newcommand{\be}{\begin{equation}}
\newcommand{\ee}{\end{equation}}
\newcommand{\bea}{\begin{eqnarray}}
\newcommand{\eea}{\end{eqnarray}}

\newcommand{\tr}{\mbox{tr}}

\newcommand{\ket}[1]{\mbox{$| #1 \rangle$}}
\newcommand{\braket}[2]{\mbox{$\langle #1  | #2 \rangle$}}
\newcommand{\proj}[1]{\mbox{$|#1\rangle \!\langle #1 |$}}

\def\tr{ \mbox{tr}}

\begin{document}

\title{
Matrix product decomposition and classical simulation of quantum dynamics \\in the presence of a symmetry}
\author{S. Singh}
\affiliation{School of Physical Sciences, the University of
Queensland, QLD 4072, Australia}
\author{H.-Q. Zhou}
\affiliation{Department of Physics, Chongqing University, Chongqing 400044, The People's 
Republic of China
}
\author{G. Vidal}
\affiliation{School of Physical Sciences, the University of
Queensland, QLD 4072, Australia}
\date{\today}

\begin{abstract}
We propose a refined {\em matrix product state} representation for many-body quantum states that are invariant under SU(2) transformations, and indicate how to extend the {\em time-evolving block decimation} (TEBD) algorithm in order to simulate time evolution in an SU(2) invariant system. The resulting algorithm is tested in a critical quantum spin chain and shown to be significantly more efficient than the standard TEBD.
\end{abstract}

%\pacs{03.67.-a, 03.65.Ud, 03.67.Hk}

\maketitle

% INTRODUCTION

Quantum many-body systems are described by a large Hilbert space, one whose dimension grows exponentially with the system's size. This makes the numerical study of {\em generic} quantum many-body phenomena computationally {\em hard}. However, quantum systems are governed by Hamiltonians made of local interactions, that is, by highly non-generic operators. As a result, physically relevant states are {\em atypical} vectors in the Hilbert space and may sometimes be described {\em efficiently}.
Systems in one spatial dimension offer a prominent example. Here the geometry of local interactions induces an anomalously small amount of bipartite correlations and an efficient representation is often possible in terms of a trial wave function known as {\em matrix product state} (MPS) \cite{MPS,MPS_review}. This, in turn, underlies the success of the {\em density matrix renormalization group} (DMRG) \cite{DMRG}, an algorithm to compute ground states, and of several recent extensions \cite{TEBD,TEBD2,other}, including the {\em time-evolving block decimation} (TEBD) algorithm to simulate time evolution \cite{TEBD}.

% WHAT WE DO

Symmetries, of fundamental importance in Physics, require a special treatment in numerical studies. Unless explicitly preserved at the algorithmic level, they are bound to be destroyed by the accumulation of small errors, in which case significant features of the system might be concealed. On the other hand, when properly handled, the presence of a symmetry can be exploited to reduce simulation costs. Whereas the latter has long been realised in the context of DMRG \cite{DMRG,DMRGsym}, the subject remains mostly unexplored for the TEBD algorithm \cite{u1}.

In this letter we undertake the study of how to enhance the MPS representation and the TEBD algorithm in systems that are invariant under the action of a Lie group ${\cal G}$.
We present an explicit theoretical construction of a refined MPS representation with built-in symmetry, and put forward a significantly faster TEBD algorithm that both preserves and exploits the symmetry. For simplicity and concreteness, we analyse the smallest non-abelian case, the SU(2) group, which is extremely relevant in the context of isotropic quantum spin systems. The analysis of the SU(2) group already contains the major ingredients of a generic group ${\cal G}$ ---in contrast with the case of an abelian U(1) symmetry \cite{u1}. In addition, it can be cast in the language of spin operators, more familiar to physicists than group representation theory. As a test, we have computed the ground state of the spin-1/2 antiferromagnetic heisenberg chain, obtaining remarkably precise two-point correlators both for short and long distances.

In preparation to describe the SU(2) MPS, we start by introducing a convenient vector basis and discuss a bipartite decomposition of states invariant under SU(2).

%%%%%%%%%%%%%%%%%%%%%%%%%%%%%%%%%%%%%%
% Total spin basis
%%%%%%%%%%%%%%%%%%%%%%%%%%%%%%%%%%%%%%

{\bf Total spin basis.--} Let $V$ be a vector space on which $SU(2)$ acts unitarily by means of transformations $e^{i\vec{v}\cdot\vec{S}}$, where matrices $S_x$, $S_y$ and $S_z$ close the Lie algebra su(2), namely $[S_{\alpha}, S_{\beta}] = i\epsilon_{\alpha\beta\gamma}S_{\gamma}$, and $\vec{v}\in {\mathbb R}^3$. A total spin basis (TSB) $\ket{^{[V]}_{jtm}}\in V$ satisfies the eigenvalue relations
\begin{eqnarray}\label{eq:eigen}
	{\vec{S}}^2\ket{^{[V]}_{jtm}}= j(j+1)\ket{^{[V]}_{jtm}}, ~~~~S_z \ket{^{[V]}_{jtm}} = m \ket{^{[V]}_{jtm}},
\end{eqnarray}
and is associated with the direct sum decomposition of $V$ into irreducible representations (irreps) of SU(2) \cite{irreps},
\begin{equation}\label{eq:direct_sum}
	V \cong \bigoplus_{j} \left(\tilde{V}^{(j)} \otimes V^{(j)}\right).
\end{equation}
Here $\tilde{V}^{(j)}$ is a $d_j$-dimensional space that accounts for the {\em degeneracy} of the spin-$j$ irrep and has basis $\ket{^{[V]}_{jt}} \in \tilde{V}^{(j)}$, where $t=1,\cdots\!, d_j$, whereas $V^{(j)}$ is a $(2j + 1)$-dimensional space that accommodates a spin-$j$ irrep and has basis $\ket{^{[V]}_{jm}} \in V^{(j)}$, where $m$ is the projection of the spin in the $z$ direction, $m =-j,\cdots\!, j$.
Each vector of the TSB factorizes into degeneracy and irrep parts as $\ket{^{[V]}_{jtm}} = \ket{^{[V]}_{jt}}\ket{^{[V]}_{jm}}$, where Eq. (\ref{eq:eigen}) only determines $\ket{^{[V]}_{jm}}$.

%%%%%%%%%%%%%%%%%%%%%%%%%%%%%%%%%%%%%%
% Bipartite Decomposition
%%%%%%%%%%%%%%%%%%%%%%%%%%%%%%%%%%%%%%

{\bf Bipartite decomposition.--} A pure state $\ket{\Psi}$ of a bipartite system with vector space $A\otimes B$ can always be expressed in terms of a TSB for $A$ and a TSB for $B$ as
\begin{equation} \label{eq:ket}
\ket{\Psi} = \sum_{j_1t_1m_1}\sum_{j_2t_2m_2} N^{j_1t_1m_1}_{j_2t_2m_2} ~~ \ket{^{[A]}_{j_1m_1t_1}}~\ket{^{[B]}_{j_2t_2m_2}}.
\end{equation}
When $\ket{\Psi}$ is an SU(2) {\em singlet}, that is, invariant under transformations acting simultaneously on $A$ and $B$, or
\begin{equation}
(\vec{S}^{[A]}+\vec{S}^{[B]})^2\ket{\Psi}= 0, ~~~~(S_z^{[A]} + S_z^{[B]}) \ket{\Psi} = 0,
\end{equation}
then the symmetry materialises in constraints for the tensor of coefficients $N$, which splits into degeneracy and irrep parts according to \cite{Sukhi}
\begin{equation}\label{eq:singlet}
\ket{\Psi}\! =\!\sum_{j} \!\left(\sum_{t_1t_2} T^{j}_{t_1t_2} \ket{^{[A]}_{jt_1}} \ket{^{[B]}_{jt_2}} \right)\!\!\left( \sum_{m} \omega^j_{m} \ket{^{[A]}_{jm}} \ket{^{[B]}_{j-m}}\right), 
\end{equation}
where $\omega$ is completely determined in terms of Clebsch-Gordan coefficients $\braket{j_1j_2m_1m_2}{j_1j_2;jm}$ \cite{Clebsch}, namely
\begin{eqnarray}
	&&\omega^j_m \equiv \left\{ \begin{array}{ll}
	 (2j+1)^{-1/2} ~~~~~~~~~~~~~ j=0,1,2,\dots\\
	 (-1)^{m}(2j+1)^{-1/2} ~~~~ j=\frac{1}{2},\frac{3}{2},\frac{5}{2},\dots        \end{array} \right., \label{eq:omega}\\
	&&\braket{j_1j_2m_1m_2}{j_1j_2;00} = \delta_{j_1,j_2}\delta_{m_1,-m_2} \omega^{j_1}_{m_1}.
\end{eqnarray}
Eq. (\ref{eq:singlet}) is quite sensible: it says that a coefficient $N^{j_1t_1m_1}_{j_2t_2m_2}$ in Eq. (\ref{eq:ket}) may be non-zero only if ($i$) $j_1=j_2$ (only the product of two spin $j$ irreps can give rise to a spin $0$ irrep, that is, the singlet $\ket{\Psi}$) and ($ii$) $m_1=-m_2$, which guarantees that the z-component of the spin vanishes. In addition, Eq. (\ref{eq:singlet}) embodies the essence of our strategy: to isolate the degrees of freedom that are not determined by the symmetry -- in this case the degeneracy tensor $T^{j}_{t_1t_2}$. We now consider the singular value decomposition 
\begin{equation}\label{eq:SVD}
T^{j}_{t_1t_2} = \sum_t(R^{j})_{t_1 t} (\eta^{j})_t (S^{j})_{tt_2}
\end{equation}
of tensor $T^{j}_{t_1t_2}$ for a fixed $j$, and define
\begin{equation}\label{eq:Schmidt_basis}
\ket{\Psi^{[A]}_{jt}} \equiv \sum_{t_1} R_{t_1t} \ket{^{[A]}_{jt_1}}, ~~~~~ \ket{\Psi^{[B]}_{jt}} \equiv \sum_{t_2} S_{t t_2} \ket{^{[B]}_{jt_2}}.
\end{equation}
By combining Eqs. (\ref{eq:singlet}), (\ref{eq:SVD}) and (\ref{eq:Schmidt_basis}) we arrive to our {\em canonical symmetric bipartite decomposition} (CSBD)
\begin{equation}
\ket{\Psi}\! = \!\sum_{j} \!\left( \!\sum_t \eta^{j}_t \ket{\Psi^{[A]}_{jt}}\ket{\Psi^{[B]}_{jt}} \!\right)\!\!\left( \sum_{m} \omega^j_{m} \ket{^{[A]}_{jm}} \ket{^{[B]}_{j-m}}\!\right), \label{eq:CSBD}
\end{equation}
which is related to the Schmidt decomposition
\begin{equation}
	\ket{\Psi} = \sum_{\alpha} \lambda_{\alpha} \ket{\Phi^{[A]}_{\alpha}}  \ket{\Phi^{[B]}_{\alpha}}, \label{eq:Schmidt}
\end{equation}
by the identifications $\alpha \rightarrow (jtm)$, $~\lambda_{\alpha} \rightarrow \eta^{j}_{t} \omega^{j}_{m}~$ and
\begin{eqnarray}
	\ket{\Phi^{[A]}_{\alpha}} \rightarrow \ket{\Psi^{[A]}_{jt}} \ket{^{[A]}_{jm}},
	~~~\ket{\Phi^{[B]}_{\alpha}} \rightarrow \ket{\Psi^{[B]}_{jt}} \ket{^{[B]}_{j-m}},
\end{eqnarray}
where some of the Schmidt coefficients $\lambda_{\alpha}$ are negative.
%%%%%%%%%%%%%%%%%%%%%%
\begin{figure}
  \includegraphics{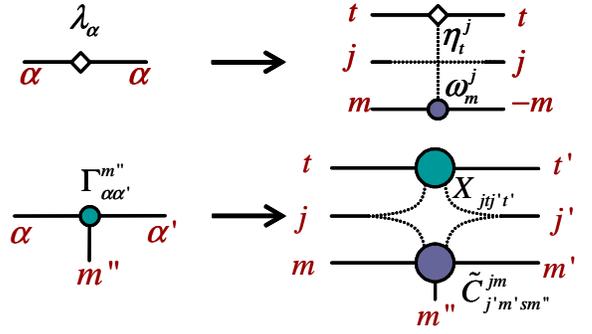}
\caption{Diagramatic representation of tensors $\lambda$ and $\Gamma$ of an MPS and tensors $(\eta,\omega)$ and $(X,\tilde{C})$ of an SU(2) MPS.}
\label{fig:SU2_MPS1}
\end{figure}
%%%%%%%%%%%%%%%%%%%%%%%%
More generally, a state $\ket{^{[CD]}_{jtm}}$ of a bipartite system $C\otimes D$ can be expressed in terms of TSBs for $C$ and $D$ as \cite{Sukhi}
\begin{eqnarray}
	\ket{^{[CD]}_{jtm}} = \sum_{jj_1j_2} \left( \sum_{tt_1t_2} X^{jt}_{j_1t_1j_2t_2}	\ket{^{[C]}_{j_1t_1}}\ket{^{[D]}_{j_2t_2}} \right) \nonumber \\
	\left( \sum_{m m_1m_2}C^{jm}_{j_1m_1j_2m_2}\ket{^{[C]}_{j_1m_1}}\ket{^{[D]}_{j_2m_2}} \right)\label{eq:TBSs}
\end{eqnarray}
where tensor $X$ relates degeneracy degrees of freedom and tensor $C$ is given by the Clebsh-Gordan coefficients
\begin{equation} \label{eq:Clebsch-Gordan}
	C^{jm}_{j_1m_1j_2m_2} = \braket{j_1j_2m_1m_2}{j_1j_2;jm}.
\end{equation}

%%%%%%%%%%%%%%%%%%%%%%%%%%%%%%%%%%%%%%
% Matrix Product Decomposition
%%%%%%%%%%%%%%%%%%%%%%%%%%%%%%%%%%%%%%

{\bf Matrix Product decomposition.--} We now consider a chain of $n$ quantum spins with spin $s$, represented by a 1D lattice where each site, labelled by $r$ ($r=1, \dots, n$), carries a $(2s+1)$-dimensional irrep of SU(2). The coefficients $c_{m_1 m_2 \dots m_n}$ of a state $\ket{\Psi}$ of the lattice,
\begin{equation}
	\ket{\Psi} = \sum_{m_1=1}^{2s+1} \cdots \sum_{m_n=1}^{2s+1} c_{m_1 m_2 \dots m_n} \ket{_{m_1}^{[1]}}\ket{_{m_2}^{[2]}}\cdots\ket{_{m_n}^{[n]}},
\end{equation}
where $\{\ket{^{[r]}_m}\}$ is a basis for site $r$ with $S_z^{[r]}\ket{^{[r]}_m} = m\ket{^{[r]}_{m}}$, can be codified as an MPS \cite{MPS,MPS_review},
\begin{equation}
c_{m_1 \dots m_n} = \sum_{\alpha_1\cdots \alpha_{n-1}} \Gamma^{[1]m_1}_{\alpha_1} \lambda^{[1]}_{\alpha_1} \Gamma^{[2]m_2}_{\alpha_1\alpha_2}\lambda^{[2]}_{\alpha_2}\cdots\Gamma^{[n]m_n}_{\alpha_{n-1}}.
\label{eq:mps}
\end{equation}
Following the conventions of \cite{TEBD}, here $\lambda^{[r]}_{\alpha}$ are the Schmidt coefficients of $\ket{\Psi}$ according to the bipartition $[1\cdots r]:[r\!+\!1\cdots n]$ of the spin chain, while tensor $\Gamma^{[r]m}_{\alpha\beta}$ relates Schmidt vectors for consecutive bipartitions,
\begin{equation}\label{eq:Gamma}
\ket{\Phi_{\alpha}^{[r\cdots n]}} = \sum_{m=1}^{2s+1} \Gamma^{[r]m}_{\alpha\beta} \lambda^{[r]}_{\beta} ~ \ket{^{[r]}_m}\ket{\Phi_{\beta}^{[r\!+\!1\cdots n]}}.
\end{equation}
When $\ket{\Psi}$ is a singlet, that is
\begin{equation}
(\sum_{r}\vec{S}^{[r]})^2\ket{\Psi}= 0, ~~~~\sum_r S_z^{[r]} \ket{\Psi} = 0,
\end{equation}
then Eqs. (\ref{eq:CSBD}) and (\ref{eq:TBSs}) supersede Eqs. (\ref{eq:Schmidt}) and (\ref{eq:Gamma}) and each tensor $\lambda$ and $\Gamma$ in Eq. (\ref{eq:mps}) decomposes into {\em degeneracy} and {\em irrep} parts, see Fig. (\ref{fig:SU2_MPS1}),
\begin{eqnarray} 
	\lambda_{\alpha} = \lambda_{(jtm)} &\rightarrow& \eta^{j}_{t} ~\omega^{j}_{m}, \label{eq:SU2_MPS1}\\ 
	\Gamma_{\alpha\alpha'}^{m''}=\Gamma_{(jtm)(j't'm')}^{(s m'')} &\rightarrow& X_{jtj't'}~\tilde{C}^{jm}_{j'm'sm''}, \label{eq:SU2_MPS2}
\end{eqnarray}
where $\tilde{C}$ is related to Clebsch-Gordan coefficients $C$ by
\begin{equation}
	\tilde{C}^{jm}_{j'm'j''m''} \equiv (-1)^{2j'}(\omega^{j'}_{m'})^{-1} C^{jm}_{j'm'j''m''}.
\end{equation}
The SU(2) MPS is defined through Eqs. (\ref{eq:SU2_MPS1})-(\ref{eq:SU2_MPS2}). 
In this representation, the constraints imposed by the symmetry are used to our advantage. By splitting tensors $\lambda$ and $\Gamma$, we achieve two goals simultaneously. 
On the one hand, the resulting MPS is guaranteed, by construction, to be invariant under SU(2) transformations. That is, any algorithm based on this representation will preserve the symmetry exactly and permanently. On the other hand, all the degrees of freedom of $\ket{\Psi}$ are concentrated in smaller tensors $\eta$ and $X$ (tensors $\omega$ and $\tilde{C}$ are specified by the symmetry), and thus the SU(2) MPS is a more economical representation. If $|\cdot|$ denotes the number of coefficients of a tensor, then
\begin{eqnarray}
	|\lambda| \equiv \sum_j (2j+1) d_j &\rightarrow& |\eta|\equiv \sum_j d_j, \label{eq:space1}\\
	|\Gamma| = (2s+1)|\lambda||\lambda'| &\rightarrow& |X| \equiv \sum_{(j,j')} d_{j}d_{j'}, \label{eq:space2}
\end{eqnarray}
where $\lambda$ and $\lambda'$ are the tensors to the left and to the right of $\Gamma$, and where, following spin composition rules, the last sum is restricted to pairs $(j,j')$ such that $|j - j'| \leq s$.
%%%%%%%%%%%%%%%%%%%%%%%%%%%%%%%%%%%%%%%%%%%%
\begin{figure}
  \includegraphics[width=8.5cm]{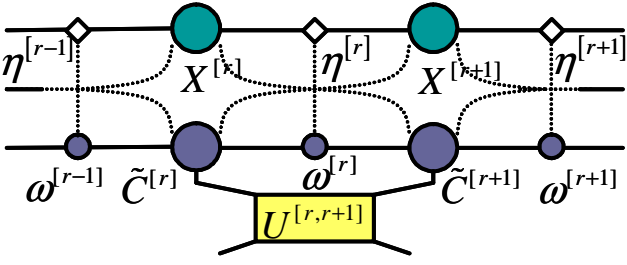}\\
\caption{The TEBD algorithm is based on updating the MPS when a gate $U$ acts on two neighboring sites. This diagramm generalizes Fig. (3.$i$) in \cite{iTEBD} after the replacements $\lambda \rightarrow (\eta,\omega)$ and $\Gamma\rightarrow (X,\tilde{C})$ of Eqs. (\ref{eq:SU2_MPS1})-(\ref{eq:SU2_MPS2}) for an SU(2) MPS. }
\label{fig:SU2_MPS2}
\end{figure}
%%%%%%%%%%%%%%%%%%%%%%%%%%%%%%%%%%%%%%%%%%%%%

\begin{figure}
  \includegraphics[width=8.5cm]{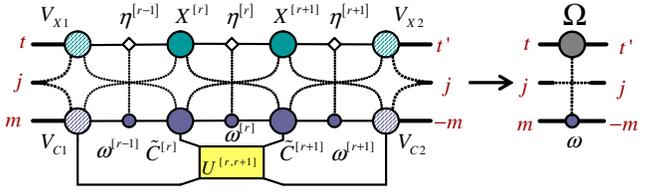}\\
\caption{Key step of the TEBD algorithm for an SU(2) MPS, analogous to Figs. (3.$i$)-(3.$iii$) in \cite{iTEBD} for a regular MPS.---
%The TEBD algorithm for an SU(2) MPS is significantly more involved than for a regular MPS. Here we illustrate the key step, analogous to Figs. (3.$i$)-(3.$iii$) in \cite{iTEBD} for a regular MPS. 
Once $U$ has been applied on two spins, additional tensors $V_{Xi},V_{Ci}$ implement a unitary transformation required to reabsorb these spins into blocks and obtain an updated representation for the bipartition $[1\cdots r]:[r+1\cdots n]$. Then, for each fixed value of the $j$ indices (discontinuous lines), the $\eta$'s, $X$'s and $V_X$'s are multiplied together and the result, with a weight coming from the product of the $\omega$'s, $\tilde{C}$'s, $V_C$'s and $U$ [that can be pre-computed because none of these tensors depend on $\ket{\Psi}$], is added together to give rise to tensor $\Omega$. A singular value decomposition of $\Omega_{jtjt'}$ for each value of $j$ ensues, see Eq. (\ref{eq:time2}), and minor rearrengements finally lead to updated tensors $X'^{[r]}$, $\eta'^{[r]}$ and $X'^{[r+1]}$.
}
\label{fig:SU2_MPS3}
\end{figure}

{\bf Simulation of time evolution.--} Our next step is to generalize the TEBD algorithm \cite{TEBD} to the simulation of SU(2)-invariant time evolution. This reduces to explaining how to update the SU(2) MPS when an SU(2)-invariant gate $U$ acts between contiguous sites, see Fig. (\ref{fig:SU2_MPS2}). The update is achieved by following steps analogous to those of the regular TEBD algorithm, see Fig. (3) of \cite{iTEBD}, involving tensor multiplications and one singular value decomposition (SVD), Fig. (\ref{fig:SU2_MPS3}). However, all these manipulations involve now smaller tensors, and only tensors $X$ and $\eta$ of the SU(2) MPS need to be updated. This results in a substantial reduction of computational space and time, and thus an increase in performance. For instace, the SVD of $\Theta$ in Fig. (3) of \cite{iTEBD}, where $|\Theta| \approx (2s+1)^2|\lambda|^2$, is now replaced with the SVD of $\Omega_{jtjt'}$
(see Fig. (\ref{fig:SU2_MPS3})) 
for each value of $j$, where $|\Omega_{jtjt'}| = (\sum_{j'\geq j-2s}^{j+2s} d_{j'})^2$. The cost $c_{svd}(A)$ of computing the SVD of a matrix $A$ grows roughly as $|A|^{3/2}$ and is the most expensive manipulation of the TEBD algorithm. We obtain the following comparative costs:

\begin{eqnarray}
c_{svd}(\Theta) &\sim&  \left((2s+1)\sum_j [(2j+1)d_j] \right)^3,\label{eq:time1}\\ 
 c_{svd}(\Omega) &\sim& \sum_j \left(\sum_{j'\geq j-2s}^{j+2s} d_{j'}\right)^3.
\label{eq:time2}
\end{eqnarray}

%%%%%%%%%%%%%%%%%%%%%%%%%%%%%%%%%%%%%%%%%%%%
%%%%%%%%%%%%%%%%%%%%%%%%%%%%%%%%%%%%%%%%%%%%%
 
{\bf Example.--} For illustrative purposes, we consider a quantum spin chain with $s=1/2$ and with Hamiltonian,
\begin{equation}
	H = \sum_{r} ( S_x^{[r]}S_x^{[r+1]} + S_y^{[r]}S_y^{[r+1]} + S_z^{[r]}S_z^{[r+1]}),
\label{eq:heisenberg}
\end{equation}
that is, the spin-1/2 antiferromagnetic Heisenberg model, which is SU(2) invariant and quantum critical at zero temperature. We have computed an SU(2) MPS approximation to the ground state of $H$, in the limit $n\rightarrow \infty$ of an infinite chain, by simulating imaginary-time evolution \cite{iTEBD} starting from a state made of nearest-neighbor singlets $(\ket{^{[r]}_{1/2}}\ket{^{[r+1]}_{-1/2}}-\ket{^{[r]}_{-1/2}}\ket{^{[r+1]}_{1/2}})/\sqrt{2}$. With the constraint $\sum_{j}d_j=600$, we have obtained that the following irreps $j$, with degeneracies $d_j$, contribute to the odd and even bipartitions \cite{even_odd} of the resulting state,
\begin{equation}
\begin{tabular}{|c|c|c|c|c|c|}
  \hline
  $j$ & 0 & 1 & 2 & 3 & 4 \\ \hline
  $d_{j}$ & 117 & 247 & 176 & 55 & 5 \\
  \hline
\end{tabular} ~~~
\begin{tabular}{|c|c|c|c|c|c|}
  \hline
  $j$ & $\frac{1}{2}$ & $\frac{3}{2}$ & $\frac{5}{2}$ & $\frac{7}{2}$ & $\frac{9}{2}$\\ \hline
  $d_{j}$ & 220 & 242 & 115 & 22 & 1 \\
  \hline
\end{tabular} \nonumber
\end{equation}
Eqs. (\ref{eq:space1})-(\ref{eq:time2}) show substantial computational gains,
\begin{equation}
\frac{|\Gamma|}{|X|} \approx \frac{10^{7}}{2 \times 10^{5}} = 50, ~~~\frac{c_{svd}(\Theta)}{c_{svd}(\Omega)}\approx \frac{9\times 10^{10}}{3\times 10^8}= 300,
\end{equation}
that is, with a regular MPS, storing the same state would require about 50 times more computer memory, while performing each SVD would be about 300 times slower.

\begin{figure}
  \includegraphics[width=8.5cm]{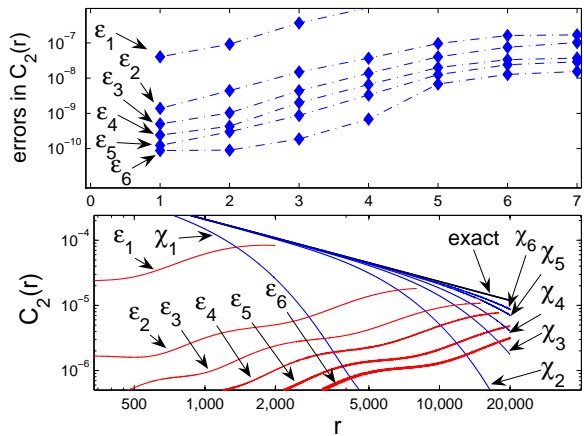}\\
\caption{
(Up) Errors in the two-point correlator $C_2(r)$ for $1\leq r \leq 7$, when using an SU(2) MPS of different sizes $\chi_i \in \{350,700,1110,1450,1800,2200\}$. 
Here $\chi$ is roughly $|\lambda|$ in Eq. (\ref{eq:space1}), that is, the rank of an equivalent (regular) MPS. The lowest line, $\chi_6=2200$, shows the errors in the data presented in the table. 
(Down) Numerical results for $C_2(r)$ for up to $r=20,000$ sites, for different sizes $\chi_i$, together with corresponding errors $\epsilon_i$.
}
\label{fig:SU2_MPS4}
\end{figure}

We have computed the two-point correlators $C^{\vartriangle}_2(r) \equiv \langle S_z^{[0]}S_z^{[r]}\rangle$ and $C^{\triangledown}_2(r) \equiv \langle S_z^{[1]}S_z^{[r+1]}\rangle$ \cite{2point_correlator}, and the average $C_2(r) \equiv (C^{\vartriangle}_2(r)+ C^{\triangledown}_2(r))/2$ \cite{translational}. For small $r$ they read:
\begin{equation}
\begin{tabular}{|c|l|l|l|}
  \hline
  $r$ & $C_2^{\vartriangle}(r)$ & $C_2^{\triangledown}(r)$ & $C_2(r)$\\ \hline
  1 & -0.14800224748 & -0.14742920605 & -0.147715726[7]\\
  2 & ~0.06067976982 & ~0.06067976991 & ~0.060679769[9] \\
  3 & -0.05037860908 & -0.05011864581 & -0.050248627[4] \\
  4 & ~0.03465277614 & ~0.03465277645 & ~0.034652776[3] \\
  5 & -0.0309785296  & -0.0308021901  & -0.03089036[0] \\
  6 & ~0.024446726   & ~0.024446726   & ~0.0244467[26] \\
  7 & -0.022565932   & -0.022430482   & -0.0224982[1]\\
  \hline
\end{tabular} \nonumber
\end{equation}
where, for $C_2(r)$, the square brakets show the first digits that differ from the exact solution \cite{exactZZ}, from which we recover {\em e.g.} 9 significant digits for $r=1$. An expression for the correlator $C_2(2)$ is also known for large $r$ \cite{Luky}. There, for $r \approx 4,000$, $10,000$ and $13,000$, our results approximate the asymptotical solution with an error of $1\%$, $5\%$ and $10\%$ respectively, see Fig. (\ref{fig:SU2_MPS4}). For comparison, with a regular MPS and similar computational resources, we lose three digits of precision for $r=1$, whereas a $10\%$ error is already achieved for $r\approx 500$ instead of $r\approx 13,000$. 

%%%%%%%%%%%%%%%%%%%%%%%%%%%%%%%%%%%%%%%%%%%%
%%%%%%%%%%%%%%%%%%%%%%%%%%%%%%%%%%%%%%%%%%%%%

{\bf Final Remarks.--} The above test with a critical spin-1/2 chain unambiguously demonstrates the superiority of the SU(2) MPS and TEBD with respect to their non-symmetric versions. Most promissingly, these techniques can now be used to address systems that remain otherwise largely unaccessible to numerical analysis due to a large dimension of the local Hilbert space. These include a chain made of large spins, say $s=4$, or a spin ladder with several legs. We regard the latter as a chain with several spins per site, where each site decomposes into SU(2) irreps as in Eq. (\ref{eq:direct_sum}) \cite{Sukhi}. 

In addition, the SU(2) MPS is not restricted to the representation of SU(2) singlets. On the one hand, it can be used to represent {\em any} SU(2) invariant mixed state $\rho$ of the chain, which decomposes as (see Eq. (\ref{eq:direct_sum}))
\begin{equation}
\rho = \bigoplus_j \rho_j \otimes I_{2j+1}
\end{equation}
This is achieved by attaching, to the end of the chain, an environment $E$ that duplicates the subspace $V$ of the chain on which $\rho$ is supported, and by considering a singlet purification $\ket{\Psi_{\rho}^{VE}}$, where $\rho=\tr_{E} \proj{\Psi_{\rho}^{VE}}$. We first build an SU(2) MPS for the purification and then we trace out $E$. The resulting structure is a matrix product representation that retains the advantages of the SU(2) MPS.  In particular, notice that when $\rho$ corresponds to one single irrep $j$, 
\begin{equation}\label{eq:rho}
\rho = \frac{1}{2j+1}\sum_{m=-j}^{m} \proj{^{[V]}_{jm}}
\end{equation}
then the environment is a site with a spin $j$, and the chain together with the environment is just an extended spin chain, with the purification being of the form
\begin{equation}\label{eq:purification}
\ket{\Psi_{\rho}} = \frac{1}{\sqrt{2j+1}}\sum_{m=-j}^j\ket{^{[V]}_{jm}}\ket{^{[E]}_{jm}}.
\end{equation}
On the other hand, the SU(2) MPS can also be modified to represent any pure state $\ket{^{[V]}_{jm}}$ of the chain with well defined $j$ and $m$. To see this, we first consider a mixed state $\rho$ as in Eq. (\ref{eq:rho}), that is, a symmetrization of $\ket{^{[V]}_{jm}}$, and then a purification $\ket{\Psi_{\rho}}$ for $\rho$ as in Eq. (\ref{eq:rho}), for which we can build an SU(2) MPS. Finally, we recall that $\ket{^{[V]}_{jm}} = \braket{^{[E]}_{jm}}{\Psi_{\rho}}$, which leads to a simple, SU(2) MPS-like representation for $\ket{^{[V]}_{jm}}$ in terms of the SU(2) MPS for the purification $\ket{\Psi_{\rho}}$. The time-evolution simulation techniques described in this paper can be applied to the above generalized representations.

Near the completion of this paper, we became aware of related results by I. McCulloch derived independently in the context of DMRG \cite{Mcculloch}.

The authors thank S. Lukyanov for helpful communications. G.V. acknowledges support from the Australian Research Council through a Federation Fellowship.


\begin{thebibliography}{99}

\bibitem{MPS} M. Fannes, B. Nachtergaele and R. F. Werner, Comm. Math. Phys. {\bf
144}, 3 (1992), pp. 443-490. S. \"Ostlund and S. Rommer, Phys. Rev.
Lett. {\bf 75}, 19 (1995), pp. 3537.

\bibitem{MPS_review} D. Perez-Garcia, F. Verstraete, M.M. Wolf, J.I. Cirac, quant-ph/0608197.

\bibitem{DMRG} S. R. White, Phys. Rev. Lett. {\bf 69}, 2863 (1992), Phys. Rev. B {\bf 48}, 10345 (1993).
U. Schollwoeck, Rev. Mod. Phys. 77, 259 (2005), cond-mat/0409292.

\bibitem{TEBD} G. Vidal, Phys. Rev. Lett. {\bf 91}, 147902 (2003),
quant-ph/0301063. G. Vidal, Phys. Rev. Lett. 93, 040502 (2004),
quant-ph/0310089.

\bibitem{TEBD2} A. J. Daley et al, J. Stat. Mech.: Theor. Exp.
(2004) P04005, cond-mat/0403313.

\bibitem{other} S. R. White and A. E. Feiguin, Phys. Rev. Lett.
93, 076401 (2004). F. Verstraete, D. Porras, J. I. Cirac Phys. Rev. Lett. 93, 227205
(2004); 

\bibitem{DMRGsym} S. Rommer, S. Ostlund, Phys. Rev. B 55, p.2164 (1997); I. P. McCulloch and M. Gul\'acsi, Europhys. Lett. {\bf 57}, 852 (2002); S. Bergkvist, I. McCulloch, A. Rosengren, cond-mat/0606265.

\bibitem{u1} The simple case of an abelian U(1) symmetry was considered by one of us in the context of bridging between DMRG and TEBD algorithms \cite{TEBD2}.

\bibitem{irreps} N. Landsman, math-ph/9807030.

\bibitem{Sukhi} S. Singh et al, {\em in preparation}.

\bibitem{Clebsch} See, for instance, http://en.wikipedia.org/wiki/Clebsch-Gordan$\_$coefficients.

\bibitem{iTEBD} G. Vidal, cond-mat/0605597.

\bibitem{even_odd} For semi-integer spins, e.g. $s=\frac{1}{2}$, the CSBD of Eq. (\ref{eq:CSBD}) for a partition $[1\cdots r]:[r\!+\!1 \cdots n]$ of the chain has only integer (semi-integer) values of $j$ when $r$ is even (odd).

\bibitem{2point_correlator} $C_2^{w}(1)$ ($w=\vartriangle,\triangledown$) are computed by contracting a small tensor network involving the tensors $(\eta,\omega)$ and $(X,\tilde{C})$ for two contiguous sites. For $r>1$, $C^w_2(r)$ is computed in the same way, but first we simulate $r-1$ (SU(2) invariant) swap gates that bring the two relevant sites together.

\bibitem{translational} We find, numerically, that the effect of \cite{even_odd} persists in the thermodynamic limit (open boundary conditions). As a result, the ground state is invariant under shifts by two (but not one) lattice sites, and $C_2^{\vartriangle}(r) \neq C_2^{\triangledown}(r)$ for odd $r$.

\bibitem{exactZZ} See J. Sato, M. Shiroishi and M. Takahashi, Nucl. Phys. B
729, 441 (2005), and references therein.

\bibitem{Luky} We use $c=1.05$ in Eq. (5.25) of S. Lukyanov, V. Terras,
% Long-distance asymptotics of spin-spin correlation functions for the XXZ spin chain, 
Nucl. Phys. B654 (2003) 323-356, hep-th/0206093.

%\bibitem{digits} The digits $x$/$y$ of $C_2(r)=xxx[yy]$ coincide with/differ from the exact solution of \cite{exactZZ}. The same number of digits are suspected to be correct for $C^{e}_2(r)$ and $C^{o}_2(r)$.

%\bibitem{TTN}  Y.Y. Shi, L.M. Duan and G. Vidal, Phys. Rev. A 74, 022320 (2006), quant-ph/0511070.

\bibitem{Mcculloch} I. McCulloch, cond-mat/0701xxx.

\end{thebibliography}
\end{document}